\documentclass[twocolumn,twoside,slac_two]{revtex4}
\usepackage{graphicx}
\usepackage{fancyhdr}
\begin{document}

\title{Using H$\alpha$ as a Tracer of the Emission Region of LS I +61 303}

%

\author{M.\ Virginia McSwain}
\affiliation{Lehigh University, Bethlehem, PA 18015, USA}

\begin{abstract}
The $\gamma$-ray binary LS I +61 303 is one of the brightest Fermi sources, with orbitally modulated emission across the electromagnetic spectrum.  Here we present H$\alpha$ spectra of LS I +61 303 that exhibit a dramatic emission burst shortly before apastron, observed as a redshifted shoulder in the line profile.  A correlated burst in radio, X-ray, and GeV emission is observed at the same orbital phase.  We interpret the source of the emission as a compact pulsar wind nebula that forms when a tidal mass stream from the Be circumstellar disk interacts with the relativistic pulsar wind.  The H$\alpha$ emission offers an important probe of the high energy emission morphology in this system.  
\end{abstract}

\maketitle

\thispagestyle{fancy}


\section{INTRODUCTION}

LS I +61 303 is a high mass X-ray binary (HMXB) that consists of an optical star with spectral type B0 Ve and an unknown compact companion in a highly eccentric, 26.5 day orbit \cite{aragona2009}.  While the system has a relatively low X-ray luminosity for a HMXB, LS I +61 303 is the 15th brightest $\gamma$-ray source included in the \textit{Fermi} LAT 1-year Point Source Catalogue (\cite{abdo2010}).  The Be disk interacts with the compact companion, producing emission that has been observed to vary with orbital phase at every wavelength across the electromagnetic spectrum, from radio to TeV (eg.\ \cite{abdo2009}, \cite{taylor1982}).  
\cite{taylor1982} found periodic radio outbursts that peak near $\phi (\rm TG) = 0.6-0.8$, and they defined the arbitrary reference for zero phase at HJD 2,443,366.775 that remains the conventional definition for LS I +61 303.  Periastron occurs at $\phi (\rm TG) = 0.275$ \cite{aragona2009}.


\section{OBSERVATIONS}

\begin{figure*}[t]
\centering
\includegraphics[width=125mm]{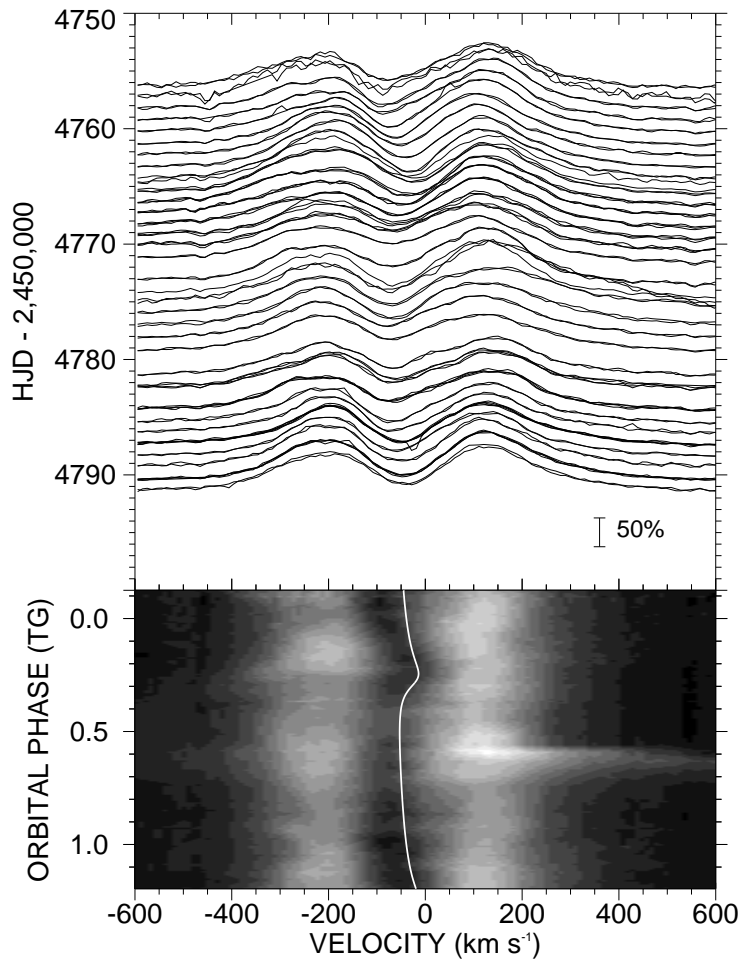}
\vspace{-.5in}
\caption{The upper plot shows the H$\alpha$ line profile over 35 continuous nights of observation at the KPNO Coud\'e Feed telescope.  The lower plot shows a gray-scale image of the same line.  Note that the lower plot of gray-scale spectra are not folded by orbital phase but are placed in the same chronological order.  From \cite{mcswain2010}.} 
\label{gray}
\end{figure*}

During 2008 October and November, we performed an intense multiwavelength observing campaign on LS I +61 303 supported by a \textit{Fermi} Cycle 1 program.  We obtained optical H$\alpha$ spectra of LS I +61 303 at the KPNO Coud\'e Feed telescope over 35 consecutive nights to study the evolution of the emission during a complete orbit \cite{aragona2009}, \cite{mcswain2010}.  The H$\alpha$ line profile exhibits a dramatic emission burst near $\phi (\rm TG) \sim 0.6$, observed as a redshifted shoulder in the line profile (see Fig.\ \ref{gray}) as the compact source moves almost directly away from the observer.   

\begin{figure*}[t]
\centering
\includegraphics[width=125mm]{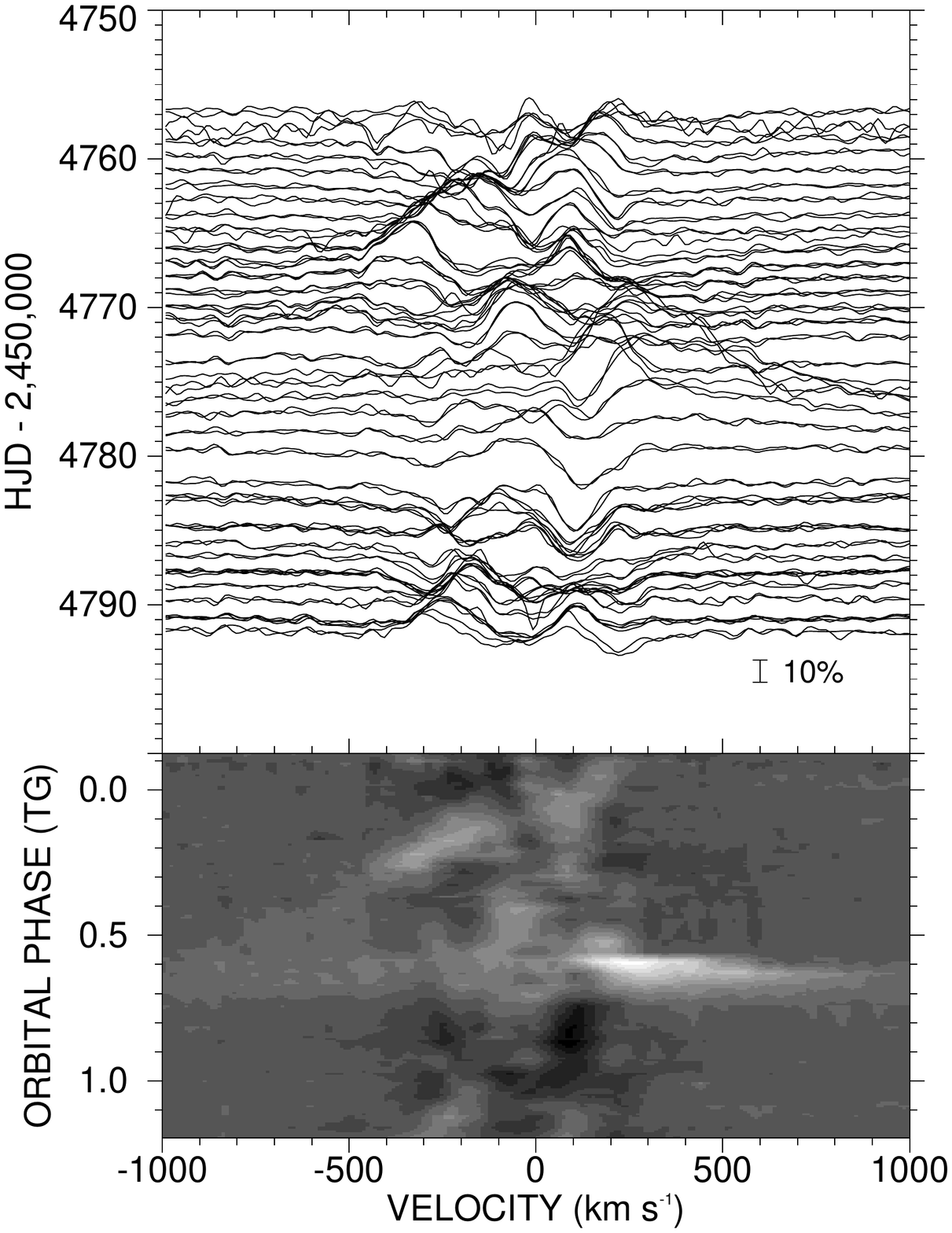}
\vspace{-.5in}
\caption{The H$\alpha$ difference spectra of LS I +61 303, shown in the same format as Figure 1.  Note the distinctive peak in emission that appears to follow a partial S-curve in the spectra.  A fast moving ``red shoulder'' of emission produces a strong redshifted feature between $\phi(\rm TG) = 0.6-0.7$.  From \cite{mcswain2010}.} 
\label{diff}
\end{figure*}

Smaller temporal changes in the red spectra suggest additional H$\alpha$ emission variability, so we subtracted the mean emission line profile to investigate the residuals carefully (see Fig.\ \ref{diff}).  During about half of the orbit, $0.9 \le \phi (\rm TG) \le 0.6$, the difference spectra reveal a partial S-shaped pattern similar to a spiral density wave that is commonly observed in Be star disks \cite{porter2003}.  \cite{zamanov1996} also observed a strong blue peak near $\phi(\rm TG) = 0.23$, which supports the development of a spiral density wave near periastron.  After this phase, the peculiar red shoulder develops.  

\begin{figure*}[t]
\centering
\includegraphics[width=135mm]{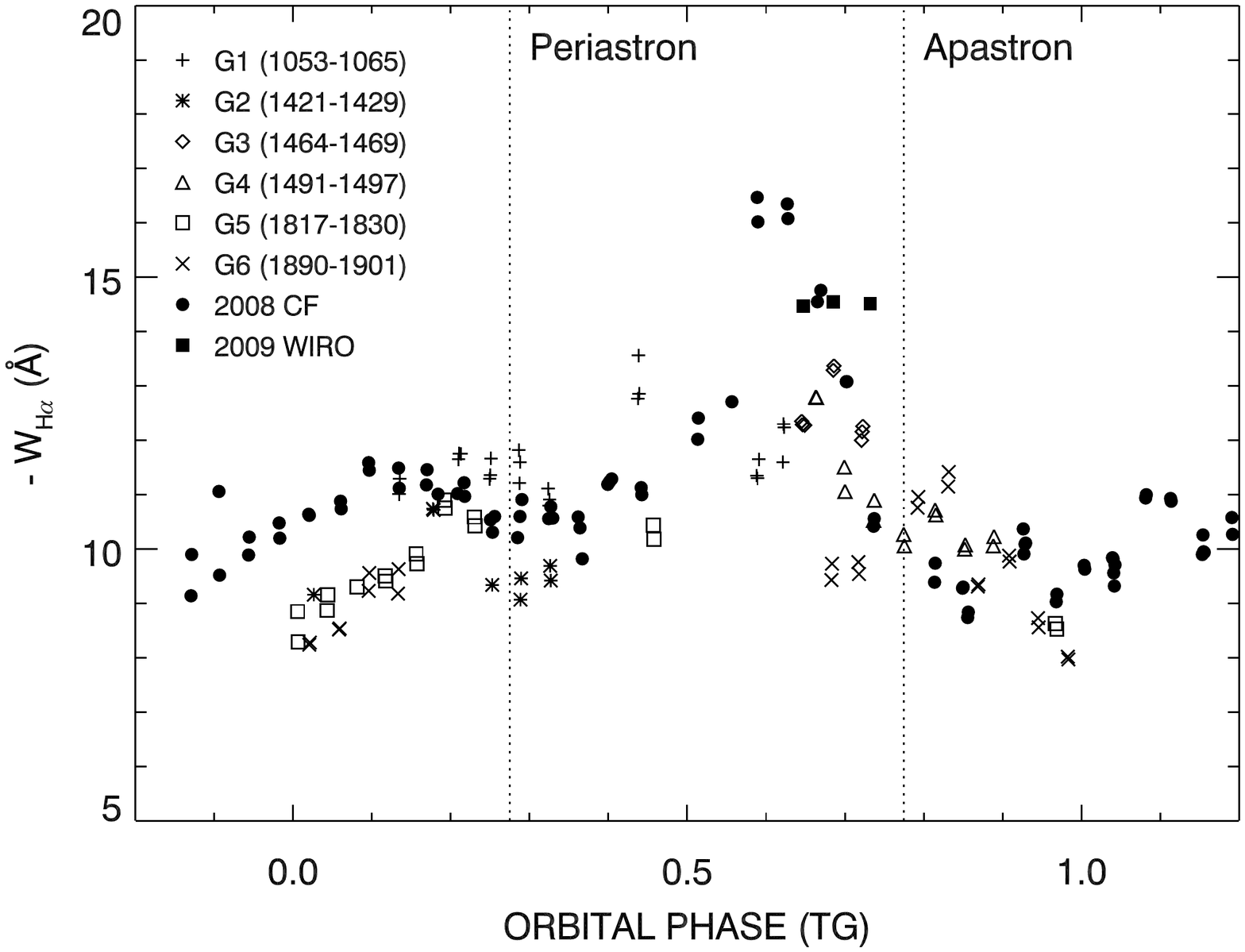}
\vspace{-.25in}
\caption{The emission strength of H$\alpha$ over multiple orbits is indicated by the equivalent width of the line ($W_{\rm H\alpha}$).  Measurements from the six runs of Grundstrom et al.\ (2007; identified by the date HJD-2,450,000) and the mean $W_{\rm H\alpha}$ from each night of our 2009 WIRO run are shown folded by orbital phase.  Note the consistent peak near $\phi(\rm TG) \sim 0.6$.  From \cite{mcswain2010}.  
} 
\label{eqwidth}
\end{figure*}

We measured the equivalent width of H$\alpha$, $W_{\rm H\alpha}$, for each spectrum by directly integrating over the line profile.  (We use the convention that $W_{\rm H\alpha}$ is negative for an emission line.)  The errors in $W_{\rm H\alpha}$ are typically about 10\% due to noise and placement of the continuum.  Figure \ref{eqwidth} shows that during our Coud\'e Feed run, $W_{\rm H\alpha}$ decreased slightly just before periastron.  Since $W_{\rm H\alpha}$ is correlated to the radius of a Be star's circumstellar disk \cite{grundstrom2006}, we interpret the decline in emission as a slight decrease in disk radius as gas is stripped away by the compact companion.  $W_{\rm H\alpha}$ then rises dramatically with the onset of the red shoulder emission component near $\phi (\rm TG) \sim 0.6$.  

Figure \ref{eqwidth} also compares our recent $W_{\rm H\alpha}$ with those measured by \cite{grundstrom2007}.  Their data were accumulated over six different observing runs over 1998--2000, and the long term differences in emission strength are substantial.  

\begin{figure*}[t]
\centering
\includegraphics[width=135mm]{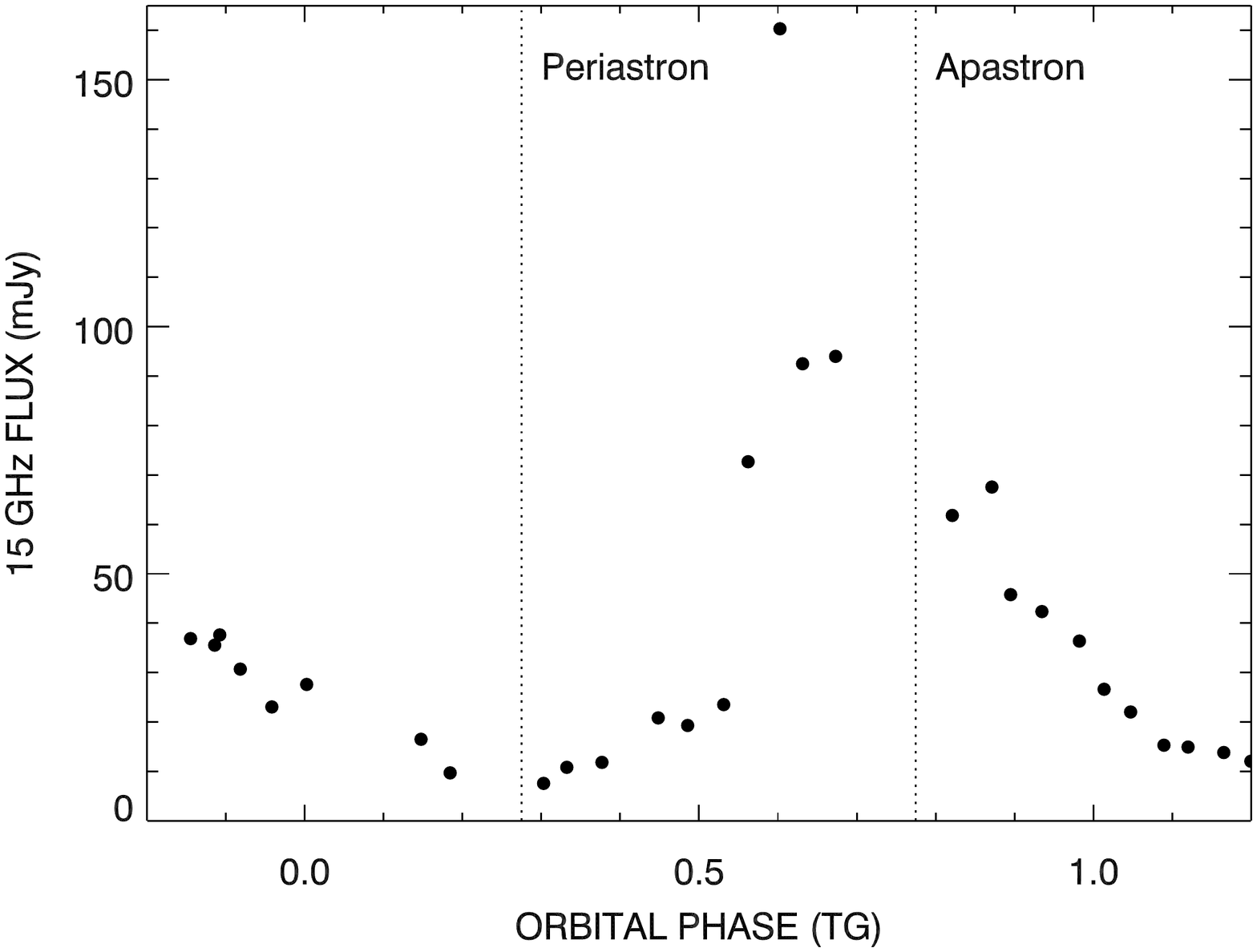}
\vspace{-.25in}
\caption{15 GHz radio flux of LS I +61 303 from the AMI array during the dates corresponding to the 2008 H$\alpha$ spectroscopy campaign.  Data courtesy of Guy Pooley.  } \label{radio}
\end{figure*}

Also during 2008 October and November, G.\ Pooley obtained nearly simultaneous radio flux coverage with the Arcminute Microkelvin Imager (AMI) array.  The 15 GHz AMI light curve (Fig.\ \ref{radio}) reveals emission that peaks at the same time as the H$\alpha$ ``red shoulder'' outburst.  
Contemporaneous \textit{RXTE} light curves from \cite{torres2010}and \textit{Fermi} light curves (\cite{dubois2011}) also reveal orbitally modulated emission that peaks just before the H$\alpha$ red shoulder, although their wide phase bins may mask a true correlation.  The H$\alpha$ emission clearly traces the high energy emission region in this system. 

\section{CONCLUSIONS}

The unusual broadness of the H$\alpha$ red shoulder emission is consistent with a Balmer-dominated shock (BDS; \cite{heng2010}).  BDS are traditionally observed around supernova remnants but are also sometimes produced within pulsar wind nebulae and other evolved stellar systems.  They form when high velocity (200--9000 km s$^{-1}$) shocks collide with the interstellar medium, manifesting themselves as optically emitting filaments.  Energetic particles and/or photons may be generated in the post-shock region of the collisionless, non-radiative shock.  Direct collisional excitation of the pre-shock atoms produces a narrow emission line component that reflects thermal conditions within the pre-shock gas.  If the energetic particles exceed the shock velocity, the pre-shock hydrogen atoms also exchange electrons with post-shock protons, manifesting themselves as broad neutral hydrogen lines  (widths $\sim 1000$ km s$^{-1}$).  The H$\alpha$ line structure in LS I +61 303 is complicated by the superposition of emission from the circumstellar disk; however, the broad red shoulder is consistent with such a BDS.  The temporary nature of the red shoulder, as well as the correlated GeV--radio emission, suggests that the BDS only forms when a high density tidal mass stream interacts with a pulsar wind in LS I +61 303.


\bigskip 
\begin{acknowledgments}

We thank Di Harmer and the staff at KPNO for their hard work to schedule and support the Coud\'e Feed observations.  Guy Pooley, Christina Aragona, Tabetha Boyajian, Amber Marsh, and Rachael Roettenbacher helped collect the data presented here and should be cheered for their heroic efforts.  This work is supported by NASA DPR numbers NNX08AV70G, NNG08E1671, NNX09AT67G, and an institutional grant from Lehigh University.  
\end{acknowledgments}


\bigskip 

\begin{thebibliography}{99} 

\bibitem{abdo2009}
Abdo A.\ A., et al.\ 2009, ApJL, 701, 123

\bibitem{abdo2010}
Abdo A.\ A., et al.\ 2010, ApJS, 188, 405

\bibitem{aragona2009}
Aragona, C., McSwain, M.~V., Grundstrom, E.~D., Marsh, A.~N., Roettenbacher, R.~M., Hessler, K.~M., Boyajian, T.~S., \& Ray, P.~S.\ 2009, ApJ, 698, 514

\bibitem{dubois2011}
Dubois, R.\ \& \textit{Fermi} LAT Collaboration, 2011, AAS Meeting 217, \#114.27

\bibitem{grundstrom2006}
Grundstrom, E.~D., \& Gies, D.~R.\ 2006, ApJL, 651, L53

\bibitem{grundstrom2007}
Grundstrom, E.\ D., Caballero-Nieves, S.\ M., Gies, D.\ R., Huang, W., McSwain, M.\ V., Rafter, S.\ E., Riddle, R.\ L., Williams, S.\ J., Wingert, D.\ W.\ 2007, ApJ, 656, 437

\bibitem{heng2010}
Heng, K.\  2010, PASA, 27, 23

\bibitem{mcswain2010}
McSwain, M.~V., Grundstrom, E.~D., Gies, D.~R., \& Ray, P.~S.\ 2010, ApJ, 724, 379

\bibitem{porter2003}
Porter, J.~M., \& Rivinius, T.\ 2003, PASP, 115, 1153 

\bibitem{taylor1982}
aylor, A.\ R., \& Gregory, P.\ C.\ 1982, ApJ, 255, 210

\bibitem{torres2010}
Torres, D.~F., et al.\ 2010, ApJL, 719, L104

\bibitem{zamanov1996}
Zamanov, R., Paredes, J.~M., Mart{\'{\i}}, J., \& Markova, N.\ 1996, Ap\&SS, 243, 269 

\end{thebibliography}

\end{document}